\documentclass{iau}
\usepackage{graphicx}
\usepackage{amsmath}
\usepackage{amssymb}
\usepackage{natbib}

\newcommand{\chan}{\textit{Chandra}}
\newcommand{\swift}{\textit{Swift}}
\newcommand{\rxte}{\textit{RXTE}}

\newcommand{\inte}{\textit{INTEGRAL}}
\newcommand{\maxi}{\textit{MAXI}}

\newcommand{\rosat}{\textit{ROSAT}}

\newcommand{\beppo}{\textit{BeppoSAX}}

\newcommand{\Msun}{\mathrm{M}_{\odot}}
\newcommand{\lum}{\mathrm{erg~s}^{-1}}
\newcommand{\flux}{\mathrm{erg~cm}^{-2}~\mathrm{s}^{-1}}
\newcommand{\fluence}{\mathrm{erg~cm}^{-2}}

\newcommand{\mdotarea}{\mathrm{g~cm}^{-2}~\mathrm{s}^{-1}}
\newcommand{\nh}{\mathrm{cm}^{-2}}

\newcommand{\sax}{SAX J1810.8--2609}

\def \mnras {\textit{MNRAS}}
\def \apj {\textit{ApJ}}
\def \apjs {\textit{ApJS}}

\def \aap {\textit{A\&A}}

\def \iaucirc {\textit{IAU Circ.}}

\title[IAUS291.~~\sax] %% short title %%
{A peculiar thermonuclear X-ray burst from the transiently accreting neutron star \\ \sax} %% full title %%

\author[N. Degenaar \& R. Wijnands]  %% short author list %%
{N. Degenaar$^{1}$\thanks{Hubble fellow}
 \and R. Wijnands$^2$}

\affiliation{$^1$University of Michigan, Dept. of Astronomy, 500
  Church St, Ann Arbor, MI 48109, USA \\
 email: {\tt degenaar@umich.edu} \\[\affilskip]
$^2$Astronomical Institute "Anton Pannekoek", University of Amsterdam\\ Postbus 94249, 1090 GE Amsterdam, The Netherlands \\email: {\tt r.a.d.wijnands@uva.nl}}

\pubyear{2012}
\volume{291}  %% insert here IAU Symposium No.
\jname{\mbox{Neutron Stars and Pulsars: Challenges and Opportunities after 80 years}}
\editors{J. van Leeuwen, ed.} 
\begin{document}

\maketitle

%% -- Abstract ----------------------------------
\begin{abstract}
We report on a thermonuclear (type-I) X-ray burst that was detected from the neutron star low-mass X-ray binary \sax\ in 2007 with \swift. This event was  longer ($\simeq$20 min) and more energetic (a radiated energy of $E_{\mathrm{b}}\simeq6.5\times10^{39}$~erg) than other X-ray bursts observed from this source. A possible explanation for the peculiar properties is that the X-ray burst occurred during the early stage of the outburst when the neutron star was relatively cold, which allows for the accumulation of a thicker layer of fuel. 
We also report on a new accretion outburst of \sax\ that was observed with \maxi\ and \swift\ in 2012. The outburst had a duration of $\simeq$17 days and reached a 2--10 keV peak luminosity of $L_{\mathrm{X}}$$\simeq$$3 \times 10^{37}~(D/\mathrm{5.7~kpc})^2~\lum$. This is a factor $>$10 more luminous than the two previous outbursts observed from the source, and classifies it as a bright rather than a faint X-ray transient.
\end{abstract}

% add below any authors, subjects and objects for indexing 
%   add more lines if necessary
%   but leave all lines commented out
%\index[author]{LastName1, Initials|textbf}
%\index[author]{LastName2, Initials|textbf}
%\index[subject]{Keyword1}
%\index[subject]{Keyword2}
%\index[object]{Object1}
%\index[object]{Object2}

\firstsection % if your document starts with a section,
\section{Introduction}
\sax\ is a transient neutron star LMXB that was discovered by \beppo\ on 1998 March 10 when it exhibited an accretion outburst \citep[][]{ubertini1998a}. \beppo\ and \rosat\ observations detected the source at a luminosity of $L_{\mathrm{X}} \simeq (0.2-1) \times 10^{36}~(D/\mathrm{5.7~kpc})^2~\lum$ (2--10 keV), and suggest that the outburst had a duration of $\gtrsim 13$~days \citep[][]{greiner1999,natalucci2000}. 

Soon after its discovery, a type-I X-ray burst was detected from \sax\ \citep[][]{cocchi1999_1810,natalucci2000}. These thermonuclear explosions occur on the surface of accreting neutron stars due to unstable burning of helium/hydrogen. The majority of observed X-ray bursts have a duration of $\simeq$10--100~s and generate a radiated energy output of $E_{\mathrm{b}}$$\simeq$$10^{39}$~erg \citep[e.g.,][]{galloway06,chelovekov2011}. Occasionally intermediately-long X-ray bursts are observed, which are more energetic ($E_{\mathrm{b}}$$\simeq$$10^{40-41}$~erg) and longer (tens of minutes) than normal X-ray bursts \citep[e.g.,][]{zand08,falanga08,linares09,degenaar2010_burst,degenaar2011_burst}.

Renewed activity was detected from \sax\ with \swift, \inte\ and \rxte\ in 2007 August \citep[][]{parsons2007,degenaar2007_1810,haymoz2007,fiocchi2009}. During this outburst \inte\ detected 17 X-ray bursts, which had an observed duration of $\simeq$10--30~s \citep[3--25 keV;][]{fiocchi2009,chelovekov2011}. The brightest event reached a bolometric peak flux of $F_{\mathrm{peak}}$$\simeq$$1 \times 10^{-7}~\flux$, implying a distance of $D$$\leq$5.7~kpc \citep[][]{fiocchi2009}.

\section{A long thermonuclear X-ray burst detected with \swift\ in 2007}\label{sec:burst}
On 2007 August 5 at 11:27:26 UT, the \swift/BAT was triggered by \sax\ \citep[trigger 287042;][]{parsons2007}. We investigated the trigger data and follow-up XRT observations, and conclude that the BAT triggered on a type-I X-ray burst. For the details on the reduction and analysis procedures we refer to \citet{degenaar2012_bursters}. We carried out exactly the same analysis for \sax. 

The BAT light curve shows a single $\simeq$10-s long peak. The average spectrum can be described by a black body model with a temperature of $kT_{\mathrm{bb}}\simeq3.0$~keV and an emitting radius of $R_{\mathrm{bb}}\simeq7$~km (Table~\ref{tab:burst_spec}), which is typical for the peak emission of X-ray bursts. We estimate a bolometric peak flux of $F_{\mathrm{peak}}\simeq7\times10^{-8}~\flux$ (0.01--100 keV), which is similar to that observed for other X-ray bursts of \sax\ \citep[][]{cocchi1999_1810,natalucci2000,fiocchi2009,chelovekov2011}.

\begin{figure}[t]
\begin{minipage}[b]{0.46\textwidth}%
 \caption{\swift/XRT light curve showing the decay of the X-ray burst detected from \sax\ on 2007 August 5 (BAT trigger 287042). The black and grey data points indicate WT and PC mode data, respectively. The dashed curve shows a fit to a power-law decay with an index of $\alpha=-1.43$, and the solid line a fit to an exponential function with a decay time of $\tau=129$~s. The dotted horizontal line indicates the persistent emission level.
 \label{fig:source}}
\end{minipage}%
\hspace{5mm}%
 \begin{minipage}[b]{0.50\textwidth}%
 \centering
 \includegraphics[width=\textwidth]{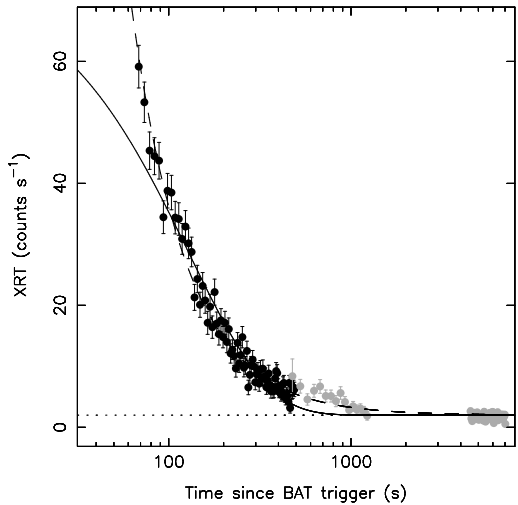}%
\end{minipage}%
\end{figure}

Automated follow-up XRT observations commenced $\simeq$65~s after the BAT trigger. The XRT light curve shows a continuous decay in count rate until it settles at a constant level $\simeq$1200~s after the BAT trigger (Figure~\ref{fig:source}). The light curve can be described by an exponential with a decay time of $\tau \simeq 129$~s, but a power law with a decay index of $\alpha \simeq -1.43$ provides a better fit (Figure~\ref{fig:source}).  The XRT data can be described by a black body model that cools along the decay (Table~\ref{tab:burst_spec}), a typical signature of X-ray bursts. 

The total estimated fluence of the X-ray burst inferred from the BAT and XRT data is $f_{\mathrm{b}}\simeq 1.6\times10^{-6}~\fluence$. For an assumed distance of $D=5.7$~kpc, this translates into a total radiated energy of $E_{\mathrm{b}} \simeq 6.5 \times 10^{39}$~erg. We use $\simeq$2~ks of XRT PC mode data obtained between $\simeq$4000--6000~s after the BAT trigger to characterize the persistent accretion emission at the time of the X-ray burst (Figure~\ref{fig:source}). This data can be described by a simple absorbed power law model with $N_{\mathrm{H}}=(6.2\pm0.1)\times10^{21}~\nh$ and $\Gamma=2.4\pm0.3$. 
We estimate a bolometric accretion luminosity of $L_{\mathrm{acc}} \simeq 5 \times 10^{35}~(D/\mathrm{5.7~kpc})^2~\lum$. The characteristics of the X-ray burst and persistent emission are summarized in Table~\ref{tab:burst_spec}.

\sax\ was covered by the \rxte/PCA Galactic bulge scan project between 1999 February 5 and 2011 October 30 \citep[][]{swank2001}, which reveals one outburst from the source (in 2007). The source was detected above the background level ($L_{\mathrm{X}} \gtrsim 3 \times 10^{35}~\lum$) between 2007 August 4 and October 28 at an average 2--10 keV luminosity of $L_{\mathrm{X}} \simeq 3 \times 10^{36}~(D/\mathrm{5.7~kpc})^2~\lum$. Non-detections on August 1 and November 1, suggest an outburst duration of $\simeq$85--92 days. 

\begin{table}
\caption{Time-resolved spectral analysis of the X-ray burst.}
\begin{center}
\begin{tabular}{p{5.7pc} p{5.7pc} p{5.7pc} p{5.7pc} p{6.7pc}}
\hline
Instrument & $\Delta t$ (s) & $kT_{\mathrm{bb}}$ (keV) & $R_{\mathrm{bb}}$ (km) & $F_{\mathrm{bol}}$ ($\flux$) \\
\hline
BAT & 0--10 & $3.0\pm 0.5$ & $7.3^{+4.9}_{-2.9}$ & $4.4\times10^{-8}$ \\
XRT/WT & 65--100 & $1.05\pm 0.04$ & $8.1\pm 0.2$ & $2.7\times10^{-9}$ \\
XRT/WT & 101--160 & $0.90\pm 0.03$ & $8.0\pm 0.2$ & $1.4\times10^{-9}$ \\
XRT/WT & 161--265 & $0.79\pm0.03$ & $7.2\pm0.2$ & $6.7\times10^{-10}$ \\
XRT/WT & 266--485 & $0.71\pm0.03$ & $6.1\pm0.2$ & $3.1\times10^{-10}$ \\
XRT/PC & 492--738 & $0.69\pm0.06$ & $5.1\pm0.4$ & $2.1\times10^{-10}$ \\
\hline
\end{tabular}
\label{tab:burst_spec}
Note. Quoted errors refer to $90\%$ confidence levels. $\Delta t$ indicates the time since the BAT trigger and $F_{\mathrm{bol}}$ the estimated bolometric flux over the interval. The simultaneous fit resulted in $N_{\mathrm{H}} =(2.2\pm0.3)\times10^{21}~\nh$ ($\chi_{\nu}^2 = 0.93$ for 281 d.o.f.) and we assumed $D=5.7$~kpc. 
\end{center}
\end{table}

\section{A new accretion outburst in 2012 seen with \maxi\ and \swift}\label{sec:2012ob}
\maxi\ monitoring observations show that \sax\ was again active between 2012 May 7--24. We estimate an average 2--20 keV luminosity of $L_{\mathrm{X}}\simeq6.3\times10^{36}~(D/\mathrm{5.7~kpc})^2~\lum$, peaking at $L_{\mathrm{X}}\simeq 2.0\times10^{37}~(D/\mathrm{5.7~kpc})^2~\lum$. The \maxi\ data suggests that the source intensity was $L_{\mathrm{X}}\gtrsim10^{36}~\lum$ for $\simeq$17~days.

A pointed \swift/XRT observation was performed on 2012 May 12 (Obs ID 32459001). The WT spectrum is best described by a combined power law and black body model with $N_{\mathrm{H}} = (0.51\pm0.02)\times10^{21}~\nh$, $\Gamma=1.67\pm0.05$, $kT_{\mathrm{bb}}=0.74\pm0.03$~keV, and $R_{\mathrm{bb}}=13.4\pm1.4$~km ($\chi_{\nu}^2 = 1.09$ for 721 d.o.f.). The resulting unabsorbed 2--10 keV model flux of $F_{\mathrm{X}} \simeq 6.9 \times 10^{-9}~\flux$ implies a luminosity of $L_{\mathrm{X}}\simeq 2.7\times10^{37}~(D/\mathrm{5.7~kpc})^2~\lum$. 

Simultaneously obtained UVOT observations using the $uw1$ filter ($\lambda_0 = 2600$~\AA) reveal an object at R.A. = 18$^{\mathrm{h}}$10$^{\mathrm{m}}$44.487$^{\mathrm{s}}$, decl. = -26$^{\circ}$09$'$01.30$''$, with an uncertainty of $0.61''$. This coincides exactly with the \chan\ position of \sax\ \citep[][]{jonker2004} and suggests that this is the UV counterpart of the LMXB. We determine (Vega) magnitudes of $uw1=18.80\pm0.12$ and $18.57\pm0.22$~mag for the two separate exposures.

\section{Discussion}\label{sec:discuss}
The X-ray burst from \sax\ observed with \swift\ is both longer and more energetic than others observed from the source \citep[][]{cocchi1999_1810,natalucci2000,fiocchi2009,chelovekov2011}. The duration is similar to that of intermediately long X-ray bursts, but the radiated energy output is an order of magnitude lower. This suggests that the X-ray burst observed from \sax\ was a normal X-ray burst, albeit with an unusual long duration. There are different explanations for such peculiar X-ray bursts \citep[see][and references therein]{degenaar2012_bursters}. 

The long X-ray burst occurred within a few days after the onset of the 2007 accretion outburst. This implies that the neutron star crust had likely not yet been significantly heated due to accretion, and hence the heat flux from the crust towards the surface was small. Combined with a low accretion rate ($\simeq$0.1\% of Eddington), this suggests that the temperature in the accreted envelop was likely low. This allows for the accumulation of a relatively thick layer of fuel before the ignition conditions are met, and may have caused its unusual properties compared to other X-ray bursts observed from the source. 

The X-ray flux observed with \maxi\ and \swift\ during the 2012 outburst of \sax\ is a factor of $>$10 higher than seen during its 1998 and 2007 outbursts. Although the previous activity of the source classified it as a faint LMXB \citep[][]{natalucci2000,jonker2004,fiocchi2009}, this demonstrates that it is actually a bright transient \citep[cf.][]{wijnands06}. It is not uncommon for bright X-ray transients to exhibit faint outbursts \citep[e.g.,][]{degenaar09_gc,degenaar2012_gc}.

\begin{table}
\caption{Characteristics of the X-ray burst and the post-burst persistent emission.}
\begin{center}
\begin{tabular}{p{19pc} p{11pc}}
\hline
Parameter & Value \\
\hline
%\multicolumn{2}{l}{\textbf{Post-burst persistent X-ray emission}} \\
Bolometric accretion luminosity, $L_{\mathrm{acc}}$ & $\simeq 4.9 \times 10^{35}~\lum$ \\
Global mass-accretion rate, $\dot{M} = R_{\mathrm{NS}}L_{\mathrm{acc}}/GM_{\mathrm{NS}}$ & $\simeq 4.1 \times 10^{-11}~\Msun~\mathrm{yr}^{-1}$ \\
Local mass-accretion rate, $\dot{m}=\dot{M}/4\pi R_{\mathrm{NS}}^2$ & $\simeq 2.1 \times 10^{2}~\mdotarea$\\
\hline
%\multicolumn{2}{l}{\textbf{X-ray burst}} \\
Bolometric X-ray burst peak flux, $F_{\mathrm{peak}}$ & $\simeq 7 \times 10^{-8}~\flux$  \\
Exponential decay time, $\tau$ & $\simeq 129$~s \\
Powerlaw decay index, $\alpha$ & $\simeq -1.43$ \\
Total duration, $t_{\mathrm{b}}$ & $\simeq 1200$~s \\
Total fluence, $f_{\mathrm{b}}$ & $\simeq 1.7 \times 10^{-6}~\fluence$ \\
Total radiated energy, $E_{\mathrm{b}}$ & $\simeq 6.5 \times 10^{39}$~erg \\
\hline
\end{tabular}
\label{tab:bursts}
Note. The quoted peak flux is unabsorbed and for the 0.01--100 keV energy range. The quoted accretion luminosity and mass-accretion rates were inferred from fitting $\simeq$2~ks of post-burst persistent emission. We assumed a distance of $D=5.7$~kpc. 
\end{center}
\end{table}

~\\
\noindent {\bf Acknowledgements}\\
ND is supported by NASA through Hubble Postdoctoral Fellowship grant number HST-HF-51287.01-A from the Space Telescope Science Institute. RW is supported by a European Research Council starting grant. This work made use of \swift\ data supplied by the UK Swift Science Data Centre at the University of Leicester, \maxi\ data provided by RIKEN, JAXA and the
MAXI team, and publicly available \rxte/PCA bulge scan light curves maintained by C. Markwardt.


\begin{thebibliography}{21}
\expandafter\ifx\csname natexlab\endcsname\relax\def\natexlab#1{#1}\fi

\bibitem[{{Chelovekov} \& {Grebenev}(2011)}]{chelovekov2011}
{Chelovekov}, I.~V. \& {Grebenev}, S.~A. 2011, \textit{Astronomy Letters}, 37, 597

\bibitem[{{Cocchi} {et~al.}(1999){Cocchi}, {Bazzano}, {Natalucci}, {Ubertini},
  {Heise}, {Muller}, {Smith}, \& {in't Zand}}]{cocchi1999_1810}
{Cocchi}, M., {Bazzano}, A., {Natalucci}, L., {et~al.} 1999, \textit{Astrophysical
  Letters and Communications}, 38, 133

\bibitem[{{Degenaar} {et~al.}(2010){Degenaar}, {Jonker}, {Torres}, {Kaur},
  {Rea}, {Israel}, {Patruno}, {Trap}, {Cackett}, {D'Avanzo}, {Lo Curto},
  {Novara}, {Krimm}, {Holland}, {de Luca}, {Esposito}, \&
  {Wijnands}}]{degenaar2010_burst}
{Degenaar}, N., {Jonker}, P.~G., {Torres}, M.~A.~P., {et~al.} 2010, \mnras,
  404, 1591

\bibitem[{{Degenaar} {et~al.}(2007){Degenaar}, {Klein-Wolt}, \&
  {Wijnands}}]{degenaar2007_1810}
{Degenaar}, N., {Klein-Wolt}, M., \& {Wijnands}, R. 2007, \textit{The Astronomer's
  Telegram}, 1175

\bibitem[{{Degenaar} {et~al.}(2012{\natexlab{a}}){Degenaar}, {Linares},
  {Altamirano}, \& {Wijnands}}]{degenaar2012_bursters}
{Degenaar}, N., {Linares}, M., {Altamirano}, D., \& {Wijnands}, R.
  2012{\natexlab{a}}, \apj, 759, 8

\bibitem[{{Degenaar} \& {Wijnands}(2009)}]{degenaar09_gc}
{Degenaar}, N. \& {Wijnands}, R. 2009, \aap, 495, 547

\bibitem[{{Degenaar} {et~al.}(2012{\natexlab{b}}){Degenaar}, {Wijnands},
  {Cackett}, {Homan}, {in't Zand}, {Kuulkers}, {Maccarone}, \& {van der
  Klis}}]{degenaar2012_gc}
{Degenaar}, N., {Wijnands}, R., {Cackett}, E.~M., {et~al.} 2012{\natexlab{b}},
  \aap, 545, A49

\bibitem[{{Degenaar} {et~al.}(2011){Degenaar}, {Wijnands}, \&
  {Kaur}}]{degenaar2011_burst}
{Degenaar}, N., {Wijnands}, R., \& {Kaur}, R. 2011, \mnras, 414, L104

\bibitem[{{Falanga} {et~al.}(2008){Falanga}, {Chenevez}, {Cumming}, {Kuulkers},
  {Trap}, \& {Goldwurm}}]{falanga08}
{Falanga}, M., {Chenevez}, J., {Cumming}, A., {et~al.} 2008, \aap, 484, 43

\bibitem[{{Fiocchi} {et~al.}(2009){Fiocchi}, {Natalucci}, {Chenevez},
  {Bazzano}, {Tarana}, {Ubertini}, {Brandt}, {Beckmann}, {Federici}, {Galis},
  \& {Hudec}}]{fiocchi2009}
{Fiocchi}, M., {Natalucci}, L., {Chenevez}, J., {et~al.} 2009, \apj, 693, 333

\bibitem[{{Galloway} {et~al.}(2008){Galloway}, {Muno}, {Hartman}, {Psaltis}, \&
  {Chakrabarty}}]{galloway06}
{Galloway}, D.~K., {Muno}, M.~P., {Hartman}, J.~M., et al.\ 2008, \apjs, 179, 360

\bibitem[{{Greiner} {et~al.}(1999){Greiner}, {Castro-Tirado}, {Boller},
  {Duerbeck}, {Covino}, {Israel}, {Linden-V{\o}rnle}, \&
  {Otazu-Porter}}]{greiner1999}
{Greiner}, J., {Castro-Tirado}, A.~J., {Boller}, T., {et~al.} 1999, \mnras,
  308, L17

\bibitem[{{Haymoz} {et~al.}(2007){Haymoz}, {Eckert}, {Shaw}, \&
  {Kuulkers}}]{haymoz2007}
{Haymoz}, P., {Eckert}, D., {Shaw}, S., \& {Kuulkers}, E. 2007, \textit{The
  Astronomer's Telegram}, 1185, 1

\bibitem[{{in 't Zand} {et~al.}(2008){in 't Zand}, {Bassa}, {Jonker}, {Keek},
  {Verbunt}, {M{\'e}ndez}, \& {Markwardt}}]{zand08}
{in 't Zand}, J.~J.~M., {Bassa}, C.~G., {Jonker}, P.~G., {et~al.} 2008, \aap,
  485, 183

\bibitem[{{Jonker} {et~al.}(2004){Jonker}, {Galloway}, {McClintock}, {Buxton},
  {Garcia}, \& {Murray}}]{jonker2004}
{Jonker}, P.~G., {Galloway}, D.~K., {McClintock}, J.~E., {et~al.} 2004, \mnras,
  354, 666

\bibitem[{{Linares} {et~al.}(2009){Linares}, {Watts}, {Wijnands}, {Soleri},
  {Degenaar}, {Curran}, {Starling}, \& {van der Klis}}]{linares09}
{Linares}, M., {Watts}, A.~L., {Wijnands}, R., {et~al.} 2009, \mnras, 392, L11

\bibitem[{{Natalucci} {et~al.}(2000){Natalucci}, {Bazzano}, {Cocchi},
  {Ubertini}, {Heise}, {Kuulkers}, {in 't Zand}, \& {Smith}}]{natalucci2000}
{Natalucci}, L., {Bazzano}, A., {Cocchi}, M., {et~al.} 2000, \apj, 536, 891

\bibitem[{{Parsons} {et~al.}(2007){Parsons}, {Barthelmy}, {Gehrels},
  {Guidorzi}, {Krimm}, {La Parola}, {Mangano}, {Markwardt}, {Marshall},
  {Palmer}, {Romano}, {Sato}, {Sbarufatti}, {Starling}, {Troja}, {vanden Berk},
  \& {Ziaeepour}}]{parsons2007}
{Parsons}, A.~M., {Barthelmy}, S.~D., {Gehrels}, N., {et~al.} 2007, \textit{GRB
  Coordinates Network}, 6706

\bibitem[{{Swank} \& {Markwardt}(2001)}]{swank2001}
{Swank}, J. \& {Markwardt}, C. 2001, in \textit{ASP Conf. Series}, Vol. 251, New Century of X-ray Astronomy, ed. {H.~Inoue \&
  H.~Kunieda}, 94

\bibitem[{{Ubertini} {et~al.}(1998){Ubertini}, {in 't Zand}, {Tesseri},
  {Ricci}, \& {Piro}}]{ubertini1998a}
{Ubertini}, P., {in 't Zand}, J., {Tesseri}, A., {Ricci}, D., \& {Piro}, L.
  1998, \iaucirc, 6838, 1

\bibitem[{{Wijnands} {et~al.}(2006){Wijnands}, {in 't Zand}, {Rupen},
  {Maccarone}, {Homan}, {Cornelisse}, {Fender}, {Grindlay}, {van der Klis},
  {Kuulkers}, {Markwardt}, {Miller-Jones}, \& {Wang}}]{wijnands06}
{Wijnands}, R., {in 't Zand}, J.~J.~M., {Rupen}, M., {et~al.} 2006, \aap, 449,
  1117

\end{thebibliography}
\end{document}